%% file: main_isbi.tex
\documentclass{article}
\usepackage{spconf,amsmath,graphicx}

\usepackage{enumitem}
\setlist{nosep, leftmargin=14pt}

\usepackage{epsfig}
\usepackage{amssymb}
\usepackage{color,soul}
\usepackage[colorlinks=true]{hyperref} 
\usepackage{pifont}
\usepackage{booktabs}  % toprule, midrule, etc
\usepackage{graphicx}
\usepackage{subfig}

\newcommand{\anonymized}[1]{#1}

\newcommand{\vmark}{\ding{51}}
\newcommand{\xmark}{\ding{55}}

\title{Predicting Knee Osteoarthritis Progression from Structural MRI using Deep Learning}
\name{\anonymized{Egor Panfilov$^{\star}$ \qquad Simo Saarakkala$^{\star, \dagger}$ \qquad Miika T. Nieminen$^{\star, \dagger}$ \qquad Aleksei Tiulpin$^{*}$}}

\address{%
$^{\star}$ \anonymized{University of Oulu, Oulu, Finland} \\
$^{\dagger}$ \anonymized{Oulu University Hospital, Oulu, Finland}}

\begin{document}
\maketitle
\begin{abstract}
Accurate prediction of knee osteoarthritis (KOA) progression from structural MRI has a potential to enhance disease understanding and support clinical trials. Prior art focused on manually designed imaging biomarkers, which may not fully exploit all disease-related information present in MRI scan. In contrast, our method learns relevant representations from raw data end-to-end using Deep Learning, and uses them for progression prediction. The method employs a 2D CNN to process the data slice-wise and aggregate the extracted features using a Transformer. Evaluated on a large cohort (n=4,866), the proposed method outperforms conventional 2D and 3D CNN-based models and achieves average precision of $0.58\pm0.03$ and ROC AUC of $0.78\pm0.01$. This paper sets a baseline on end-to-end KOA progression prediction from structural MRI. Our code is publicly available at \anonymized{\href{https://github.com/MIPT-Oulu/OAProgressionMR}{https://github.com/MIPT-Oulu/OAProgressionMR}}.
\end{abstract}
%
%%%%%%%%%%%%%%%%%%%%%%%%%%%%%%%%%%%%%%%%%%%%%%%%%%%%%%%%%%%%%%%%%%%%%%%%%%%%%%%%
\section{Introduction}\label{sec:intro}
Knee osteoarthritis (KOA) is the most common musculoskeletal disease, with estimated prevalence of more than 15\% worldwide~\cite{cui2020global}. KOA negatively affects quality of life in millions of people and brings a significant economic burden. The etiology of KOA is still under-investigated and the currently available treatment options are limited to lifestyle interventions and, at a terminal stage, total knee arthroplasty. Accurate progression prediction may advance understanding of KOA etiology, enable early disease interventions, and also support subject selection in treatment efficacy studies.

KOA is a whole joint disease, characterized by degenerative changes primarily in the knee bones and cartilage tissues, but often also in the joint capsule, muscles, etc. In clinical practice, KOA is diagnosed from demographic data, symptomatic assessments, and plain knee radiographs. The latter are used to evaluate the radiographic severity stage, quantified by the gold standard Kellgren-Lawrence grading system (KLG)~\cite{kellgren1957radiological}. Since radiography does not allow to visualize the soft tissues, such as cartilage, and is a projection imaging technique, MRI is currently actively studied in search for more comprehensive \textit{in vivo} imaging biomarkers (IB)~\cite{panfilov2021deep,morales2020learning}.

The vast majority of available IBs in KOA are manually designed to capture individual morphological and structural changes in bone (e.g. osteophytes and sclerosis), articular cartilage (e.g. partial thickness loss), ligaments, or menisci (e.g. tears). Their contribution to the disease progression is often studied separately and from small sample sizes. Deep learning (DL) enables data-driven representation learning, thus, providing an opportunity to create IBs of a complete joint by leveraging large imaging cohorts. 

Prior art on end-to-end KOA progression prediction is limited. Several works~\cite{wang2019total,tolpadi2020deep} proposed using DL to learn IBs predictive of total knee arthroplasty (TKA) within 9 and 5 years, respectively. TKA is an acknowledged structural endpoint, however, 1) it is not informative w.r.t. the stages of KOA and 2) the number of TKA cases even in the large cohorts is often prohibitively small~\cite{tolpadi2020deep}. A seminal work~\cite{tiulpin2019multimodal} introduced a finer progression criterion based on radiographic severity (i.e. KLG) change, yet studied it merely with knee radiographs. Adapting the latter definition for MRI data may result in IBs that are more specific to trajectory of radiographic KOA and potentially insightful for studying the disease etiology.

DL models for 3D medical image analysis (MIA) are typically based on Convolutional Neural Networks (CNNs). The most common architectures are planar and process volumetric images slice-by-slice, due to lower memory demand and plenty of generic data available for pre-training. Several works have specifically shown that pre-training may benefit in MIA, although, the discussion remains open~\cite{raghu2019transfusion,mustafa2021supervised}. Volumetric models have also been actively studied, particularly, for MR image analysis, where the data is inherently 3D. Since such models have significantly higher computational demand, steps have been made to improve their efficiency~\cite{kopuklu2019resource}. 

Recently, in an attempt to improve upon CNNs, a paradigm of Transformers emerged~\cite{dosovitskiy2020image}. In the context of computer vision, transformers release the strong locality prior of CNNs, thereby allowing for a richer data-driven representation learning. While Transformers show promise on large datasets, they also show higher requirements for pre-training and regularization than CNNs, especially, when used with 3D data~\cite{arnab2021vivit}. Very recently, an approach combining CNN and Transformer modules into a single architecture was shown to leverage the benefits of both paradigms~\cite{xie2021cotr,zhang2021transfuse}, and we build upon it in this paper.
To summarize, the key contributions of our work are the following:
\begin{enumerate}
    \item We propose an end-to-end method for prediction of structural osteoarthritis progression from knee MRI.
    \item We compare our hybrid CNN-Transformer with conventional 2D, (2+1)D, and 3D architectures for MRI data analysis in performance and computational efficiency.
    \item We conduct an ablation study and investigate the effect of the design choices, data reprojection, and model pre-training on the final performance.
\end{enumerate}

%%%%%%%%%%%%%%%%%%%%%%%%%%%%%%%%%%%%%%%%%%%%%%%%%%%%%%%%%%%%%%%%%%%%%%%%%%%%%%%%
\section{Method}\label{sec:method}
\subsection{Overview}
The workflow of our approach is summarized in Figure~\ref{fig1:workflow}. Here, a Dual-Echo Steady-State (DESS) MRI scan is used as an input to the predictive model. The model is trained to predict whether a radiographic KOA progression will happen in the imaged knee within the next 8 years (96 months)~\cite{tiulpin2019multimodal}.

\input{figure_1}

\subsection{Definition of Osteoarthritis Progression}
Using the approach from~\cite{tiulpin2019multimodal}, we defined the progression criteria based on the increase in disease severity measured with KLG. In KLG system, KL0 indicates no radiographic KOA (RKOA), KL1 -- doubtful, KL2 -- early, KL3 -- moderate, and KL4 -- severe RKOA~\cite{kellgren1957radiological}. We treated progression from KL0 to KL1 as no progression. Our prediction target had 3 classes: no RKOA progression within 96 months, slow RKOA progression (after 72 and within 96 months), and fast RKOA progression (within 72 months). During the evaluation, we combined the predictions for fast and slow progression classes to infer the probability of progression within 96 months. For further details, we refer the reader to~\cite{tiulpin2019multimodal}.

%%%%%%%%%%%%%%%%%%%%%%%%%%%%%%%%%%%%%%%%%%%%%%%%%%%%%%%%%%%%%%%%%%%%%%%%%%%%%%%%
\subsection{Architecture}\label{ssec:architecture}
The proposed architecture was composed of a 2D CNN encoder to extract the features slice-wise, a Transformer module to aggregate and model inter-slice feature interactions, and a final fully connected layer (\textit{2D + TRF}), as shown in Figure~\ref{fig2a}. ResNet-50 was chosen for the encoder due to its performance-efficiency trade-off~\cite{he2016deep}. In Transformer, the projection vector length was 2048, and the complete module had 4 blocks, each with 8 attention heads.

\input{figure_2}

\input{table_1}

\subsection{Reference Methods}
Several conventional 2D models were used for comparison. In (\textit{2D + FC}) (see Figure~\ref{fig2b}), the CNN output features were concatenated and then passed through two fully connected layers with ReLU non-linearity in-between. In (\textit{2D + Bi-LSTM}) (see Figure~\ref{fig2c}), a bi-directional LSTM was used instead to model the sequential inspection of slices. The ultimate states of forward and backward LSTM outputs were concatenated and fed into the classification layer.

To study the effect of scan reprojection, the proposed method was trained separately also on coronal and axial slices. Furthermore, a multi-view version of a method was implemented, where slice-wise features over all projections were used jointly in the model. Here, we studied the models with an encoder CNN shared across all views (\textit{2D$^{sh}$ + TRF}) and a individual CNN per each view (\textit{2D$^{ind}$ + TRF}). Similar approach has been previously successfully applied in knee cartilage segmentation~\cite{perslev2019one}. All compared 2D models also had ResNet-50 as an encoder CNN~\cite{he2016deep}. 

Lastly, several volumetric models (shown schematically in Figure~\ref{fig2d}) were added to the analysis: a factorized (\textit{(2+1)D})~\cite{tran2018closer}, a conventional fully volumetric (\textit{3D ResNeXt-50})~\cite{xie2016}, and one of the recently introduced efficient (\textit{3D ShuffleNet})~\cite{kopuklu2019resource}.

%%%%%%%%%%%%%%%%%%%%%%%%%%%%%%%%%%%%%%%%%%%%%%%%%%%%%%%%%%%%%%%%%%%%%%%%%%%%%%%%
\section{Experiments}\label{sec:experiments}
\subsection{Data}\label{ssec:data}
We used the data from The Osteoarthritis Initiative (OAI) dataset -- a multi-center longitudinal osteoarthritis study (\url{https://nda.nih.gov/oai/}). After data selection, we constructed a dataset comprising 2702 subjects present at the baseline examination. The knees (total of 4866) were imaged with DESS MRI protocol (3T Siemens MAGNETOM Trio scanners, quadrature T/R knee coils, sagittal view, $160$ slices, voxel size: $0.37 \times 0.37 \times 0.7mm$, matrix: $384 \times 384$, FOV: $140mm$, TR: $16.3ms$, TE: $4.7ms$, flip angle: $25^\circ$). The samples with missing BMI, KLG, or missing DESS MRI data were excluded, as well as the knees with KLG=4 or after TKA at the baseline. The final dataset demographics was: age $60.4\pm8.8$, BMI $28.3\pm4.7$, female/male - 1547/1155 subjects, KLG 0/1/2/3 - 2385/1157/894/430 knees. The number of knees in the non-progressor, fast, and slow progressor groups was 3551, 941, and 374, respectively.

To reduce the memory demand, the images were cropped to $320 \times 320 \times 128$ voxels, quantized to 8-bit intensity range, and downsampled by a factor of 2 in each dimension. In the experiments with the reprojected data, the factors were adjusted to make the in-slice resolution isotropic, while maintaining the same total number of voxels.

%%%%%%%%%%%%%%%%%%%%%%%%%%%%%%%%%%%%%%%%%%%%%%%%%%%%%%%%%%%%%%%%%%%%%%%%%%%%%%%%
\subsection{Implementation Details}\label{ssec:implementation}
The complete dataset was split into training and evaluation subsets (2019 and 683 subjects, 3607 and 1259 knees, respectively) subject-wise, maintaining similar distribution of target classes. The evaluation subset contained the data from a single hold-out institution to better assess the model generalization. The training subset was used in a 5-fold cross-validation setting to train an ensemble of prediction models. Our training budget was set to 100 epochs and we used average precision to select the best snapshot.

Data imbalance in classification problems often negatively affects model convergence~\cite{johnson2019survey}. In our experiments, the positive (i.e. progression) classes were resampled to balance the target distribution. Additionally, Focal Loss ($\gamma=2$) and weight decay of $10^{-4}$ were used to regularize the training. Adam optimizer was used, with the learning rate linearly warmed up from $10^{-5}$ for 5 epochs and then set to $10^{-4}$ for the rest of the training. The 2D models were initialized in two ways -- from pre-trained ImageNet weights and from scratch. The training subset was augmented by applying random cropping, in-slice rotation, and gamma correction.

%%%%%%%%%%%%%%%%%%%%%%%%%%%%%%%%%%%%%%%%%%%%%%%%%%%%%%%%%%%%%%%%%%%%%%%%%%%%%%%%
\subsection{Metrics}
We used average precision (AP) as our main metric and also presented ROC AUC~\cite{saito2015precision}. Based on the prevalence, AP of $0.26$ indicates random performance.
The computational efficiency of the models was measured by the number of model parameters, number of multiply-accumulate operations (MACs)~\cite{zhu2021}, and average inference time with a single sample.
The training and evaluation were done on 4 $\times$ Nvidia A100 GPUs, PyTorch version 1.7.1. The prediction time was measured with a single Nvidia  RTX 2080 Ti GPU.

\input{figure_3}

\subsection{Results}\label{ssec:results}
Our proposed method showed AP of $0.58\pm0.02$ and ROC AUC of $0.79\pm0.01$, which were the highest among the considered models (see Table \ref{tab:arch_dim}). Multiclass balanced accuracy was $0.52\pm0.02$, with the confusion matrix shown in Figure~\ref{fig3a}. The aggregation of features via Transformer yielded higher performance metrics than with \textit{FC} layer or \textit{Bi-LSTM}, although the model was larger in number of parameters. The models trained on sagittal and coronal views were equally accurate, as well as both of the multi-view variants. This finding suggests that, for the defined prediction target, even single view representations are sufficient to derive the whole scan IBs with Transformer. The volumetric models were the least accurate, with the highest AP of $0.54\pm0.03$ and ROC AUC of $0.76\pm0.01$ achieved with \textit{3D ShuffleNet} (see Figures~\ref{fig3b} and \ref{fig3c}). Interestingly, the best volumetric model was also the most efficient in terms of MACs and number of parameters, with the inference time, however, being third to the largest.

Since the investigated volumetric models were trained from scratch, contrary to the rest, an ablation study was done to understand the impact of encoder initialization on the final performance. As shown in Table \ref{tab:arch_pretrain}, using pre-trained encoders improved the performance both in AP and ROC AUC for all the 2D and multi-view models, except for (\textit{2D + Bi-LSTM}). However, even with random initialization these models performed more accurate or comparably to the volumetric ones.

\input{table_2}

%%%%%%%%%%%%%%%%%%%%%%%%%%%%%%%%%%%%%%%%%%%%%%%%%%%%%%%%%%%%%%%%%%%%%%%%%%%%%%%%
\section{Conclusion}\label{sec:conclusion}
This study introduced an end-to-end method for prediction of radiographic KOA progression from knee MRI. To our knowledge, this is the first work that investigates the end-to-end MRI-based IBs in scope of the radiographic KOA progression. We analyzed multiple design choices, and identified the top performing model configuration based on a combination of a 2D CNN and a Transformer.

In our experiments, 3D models showed the performance inferior to the one of 2D models. Future work may investigate the ways to improve regularization of 3D models and the benefits of using pre-training from large corpus of knee MRIs, e.g. via self-supervised learning~\cite{taleb2021multimodal,azizi2021big}.
While our work substantially improves over state-of-the-art, it is bounded by a single KOA progression surrogate~\cite{tiulpin2019multimodal}. Further studies should search for other informative KOA surrogate outcomes, e.g. specific to disease stages or phenotypes \cite{namiri2021deep}. While we considered structural DESS MRI  only, compositional MRI protocols (e.g. T$_2$, T$_{1\rho}$) should also be explored due to their promise in detection of early KOA changes~\cite{casula2017elevated}.
Finally, a more detailed clinically relevant analysis of the IBs learned by the model is a critical direction for follow-up research.

Our results show the high potential of DL in automatic identification of subjects who will develop degenerative KOA changes over time. We believe that the presented methodology and findings will facilitate translation of DL into the clinical management of OA.

%%%%%%%%%%%%%%%%%%%%%%%%%%%%%%%%%%%%%%%%%%%%%%%%%%%%%%%%%%%%%%%%%%%%%%%%%%%%%%%%
\section{Compliance with ethical standards}\label{sec:compliance}
This research study was conducted retrospectively using 
human subject data made available in open access by National Institutes of Health. Ethical approval was not required as confirmed by the license attached with the open access data.

%%%%%%%%%%%%%%%%%%%%%%%%%%%%%%%%%%%%%%%%%%%%%%%%%%%%%%%%%%%%%%%%%%%%%%%%%%%%%%%%
\section{Acknowledgments}\label{sec:acknowledgments}
\anonymized{The authors acknowledge the strategic funding of Infotech, University of Oulu and Finnish Center for Artificial Intelligence, and the computational resources by CSC – IT Center for Science, Finland. Huy Hoang Nguyen is kindly acknowledged for a discussion on Transformers. The authors have no relevant financial or non-financial interests to disclose.}

The OAI is a public-private partnership comprised of five contracts (N01-AR-2-2258; N01-AR-2-2259; N01-AR-2-2260; N01-AR-2-2261; N01-AR-2-2262) funded by the National Institutes of Health, a branch of the Department of Health and Human Services, and conducted by the OAI Study Investigators. Private funding partners include Merck Research Laboratories; Novartis Pharmaceuticals Corporation, GlaxoSmithKline; and Pfizer, Inc. Private sector funding for the OAI is managed by the Foundation for the National Institutes of Health. % This manuscript was prepared using an OAI public use data set and does not necessarily reflect the opinions or views of the OAI investigators, the NIH, or the private funding partners.

%%%%%%%%%%%%%%%%%%%%%%%%%%%%%%%%%%%%%%%%%%%%%%%%%%%%%%%%%%%%%%%%%%%%%%%%%%%%%%%%
\bibliographystyle{IEEEbib}
\bibliography{references_abbrev}

\end{document}

%% file: figure_1.tex
\begin{figure}[!ht]
  \centering
  \includegraphics[width=8.5cm]{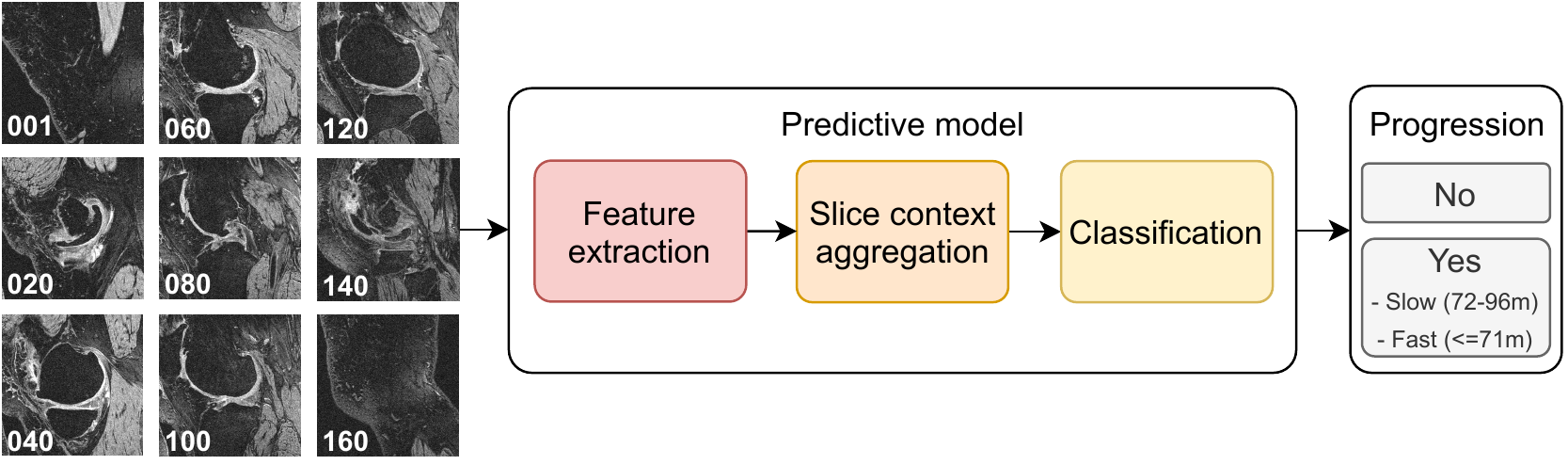}
\caption{Overview of the proposed end-to-end system for OA progression prediction from volumetric knee MRI scan. The scan is shown by evenly sampled sagittal slices}
\label{fig1:workflow}
\end{figure}

%% file: figure_2.tex
\begin{figure}[ht]
\centering\subfloat[\label{fig2a}]{\includegraphics[width=8.3cm]{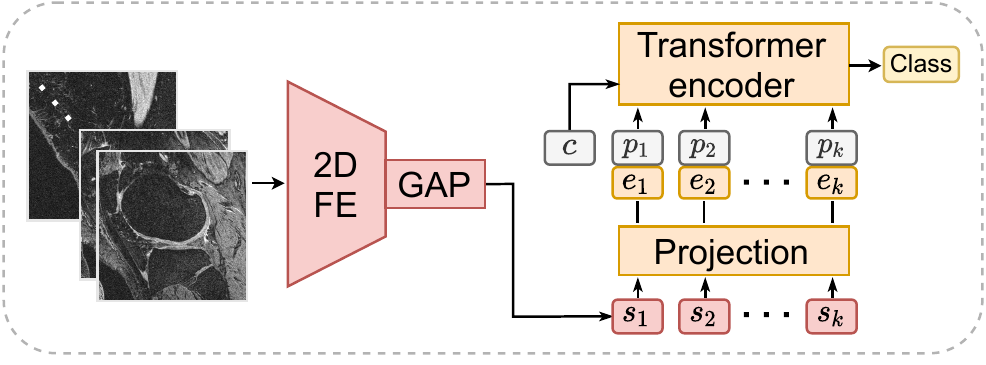}}
\vfil
\subfloat[\label{fig2b}]{\includegraphics[width=2.7cm]{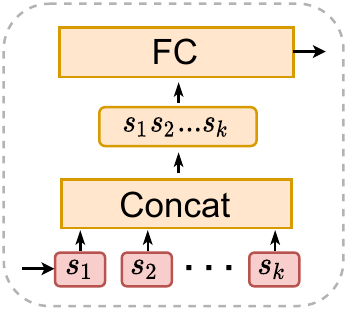}}
\hfil
\subfloat[\label{fig2c}]{\includegraphics[width=2.7cm]{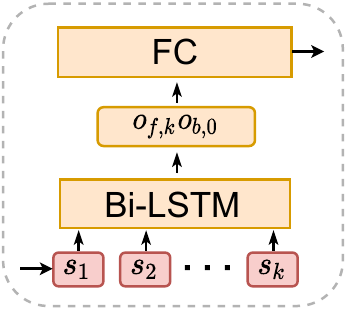}}
\hfil
\subfloat[\label{fig2d}]{\includegraphics[width=3.05cm]{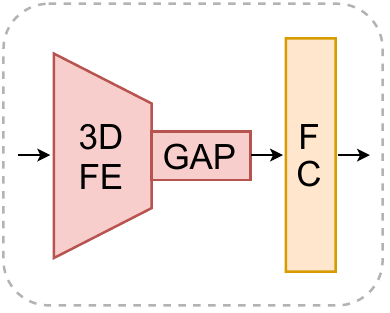}}
\caption{Schematic representation of (a) the proposed model based on 2D CNN and Transformer, (b, c) the 2D models with conventional feature aggregation via FC and Bi-LSTM, respectively, and (d) the volumetric models. In (a), slice-wise features $s_1, ..., s_k$ are projected into embeddings $e_1, ... e_k$, combined with positional $p_1, ..., p_k$ and class $c$ tokens, and processed in Transformer. $o_{f,k}$ and $o_{b,0}$ are the terminal states of Bi-LSTM output, in forward and backward passes, respectively. FE - feature extractor, GAP - global average pooling, FC - fully connected layer}
\label{fig2:archs}
\end{figure}

%% file: table_1.tex
\begin{table*}[!ht]
\centering
\resizebox{0.85\textwidth}{!}{
\begin{tabular}{l|c|c|c|c|c|c}
\toprule
\textbf{Model}          & \textbf{View} & \textbf{AP} & \textbf{ROC AUC} & \textbf{MACs, $\cdot 10^9$} & \textbf{\# params, $\cdot 10^6$} & \textbf{Inference, $ms$} \\
\midrule
Prior art (X-Ray-based) \cite{tiulpin2019multimodal} & \textit{cor proj-n} & 0.54$_{\pm 0.02}$ & 0.74$_{\pm 0.01}$ & 11 & 24 & 12.9 \\
\midrule
2D + FC               & \textit{sag} & 0.57$_{\pm 0.03}$ & 0.77$_{\pm 0.01}$      & 134                         & 91                               & 28.6                          \\
2D + Bi-LSTM             & \textit{sag} & 0.55$_{\pm 0.03}$ & 0.76$_{\pm 0.02}$      & 135                         & 29                               & 36.6                          \\
2D + TRF         & \textit{sag} & \textbf{0.58$_{\pm 0.02}$} & \textbf{0.78$_{\pm 0.01}$}      & 141                         & 133                              & 31.1                          \\
2D + TRF         & \textit{cor} & \textbf{0.58$_{\pm 0.03}$} & \textbf{0.79$_{\pm 0.01}$}      & 143                         & 133                              & 32.8                          \\
2D + TRF          & \textit{ax} & 0.55$_{\pm 0.03}$ & \textbf{0.78$_{\pm 0.01}$}      & 143                         & 133                              & 33.3                          \\
\midrule
2D$^{sh}$ + TRF   & \textit{sag}, \textit{cor}, \textit{ax} & \textbf{0.58$_{\pm 0.03}$} & \textbf{0.78$_{\pm 0.01}$}      & 443                         & 133                              & 74.8                          \\
2D$^{ind}$ + TRF   & \textit{sag}, \textit{cor}, \textit{ax} & \textbf{0.58$_{\pm 0.03}$} & \textbf{0.78$_{\pm 0.01}$}      & 443                         & 180                              & 73.3                          \\
\midrule
(2+1)D                  & \textit{sag}, \textit{cor}, \textit{ax} & 0.51$_{\pm 0.03}$ & 0.73$_{\pm 0.02}$      & 100                         & 64                               & 21.3                          \\
3D ResNeXt-50  & \textit{sag}, \textit{cor}, \textit{ax} & 0.53$_{\pm 0.03}$ & 0.75$_{\pm 0.02}$      & 21                          & 26                               & 23.6                          \\
3D ShuffleNet          & \textit{sag}, \textit{cor}, \textit{ax} & 0.54$_{\pm 0.03}$ & 0.76$_{\pm 0.01}$      & 4                           & 1                                & 45.0                          \\
\bottomrule
\end{tabular}
}
\caption{Comparison of architectures in terms of performance and computational efficiency}
\label{tab:arch_dim}
\end{table*}

%% file: figure_3.tex
\begin{figure}[ht]
\subfloat[\label{fig3a}]{\includegraphics[clip,trim=0.3cm 0.7cm 0.3cm 0.38cm,width=0.31\linewidth]{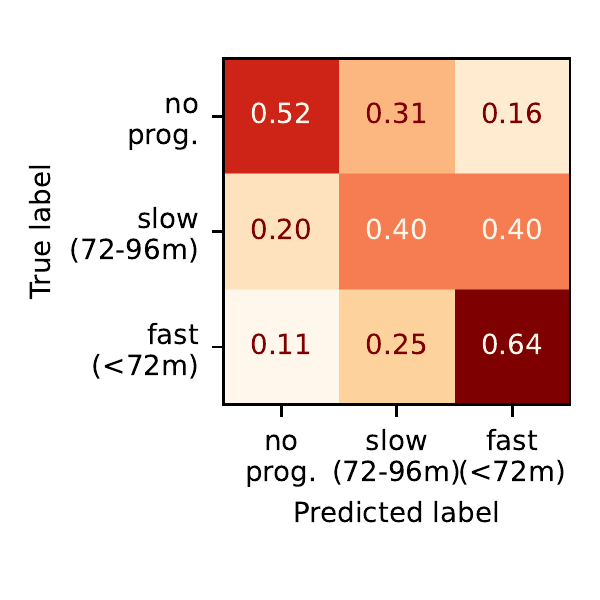}}
\hfil
\subfloat[\label{fig3b}]{\includegraphics[clip,trim=0.15cm 0.4cm 5.7cm 0.31cm,width=0.32\linewidth]{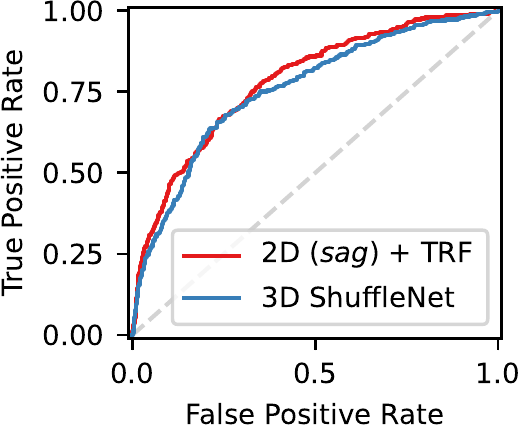}}
\hfil
\subfloat[\label{fig3c}]{\includegraphics[clip,trim=5.6cm 0.4cm 0.3cm 0.31cm,width=0.32\linewidth]{resources/figure_3bc.pdf}}
\caption{(a) Confusion matrix for the best model (before pooling the "slow" and "fast" classes). (b) ROC and (c) Precision-Recall curves for the top performing 2D and 3D models.}
\label{fig3:curves}
\end{figure}

%% file: table_2.tex
\begin{table}[!ht]
\centering
\resizebox{0.85\linewidth}{!}{
\begin{tabular}{l|c|c|c}
\toprule
\textbf{Model}        & \textbf{Init} & \textbf{AP} & \textbf{ROC AUC} \\
\midrule
2D \textit{(sag)} + FC             & \xmark                        & 0.55$_{\pm 0.02}$ & 0.76$_{\pm 0.01}$      \\
             & \vmark                        & 0.57$_{\pm 0.03}$ & 0.77$_{\pm 0.01}$      \\
\midrule
2D \textit{(sag)} + Bi-LSTM           & \xmark                        & 0.56$_{\pm 0.03}$ & 0.77$_{\pm 0.01}$      \\
           & \vmark                        & 0.55$_{\pm 0.03}$ & 0.76$_{\pm 0.02}$      \\
\midrule
2D \textit{(sag)} + TRF       & \xmark                        & 0.57$_{\pm 0.02}$ & 0.78$_{\pm 0.01}$      \\
       & \vmark                        & 0.58$_{\pm 0.02}$ & 0.78$_{\pm 0.01}$      \\
\midrule
2D$^{sh}$ \textit{(multi)} + TRF & \xmark                        & 0.55$_{\pm 0.03}$ & 0.77$_{\pm 0.01}$      \\
 & \vmark                        & 0.58$_{\pm 0.03}$ & 0.78$_{\pm 0.01}$     \\
\midrule
2D$^{ind}$ \textit{(multi)} + TRF & \xmark                        & 0.56$_{\pm 0.03}$ & 0.77$_{\pm 0.01}$      \\
 & \vmark                        & 0.58$_{\pm 0.03}$ & 0.78$_{\pm 0.01}$      \\
\bottomrule
\end{tabular}
}
\caption{Impact of CNN encoder ImageNet pre-training (\vmark) versus initialization from scratch (\xmark)}
\label{tab:arch_pretrain}
\end{table}

%% file: main_isbi.bbl
\begin{thebibliography}{10}

\bibitem{cui2020global}
A~Cui, H~Li, D~Wang, et~al.,
\newblock ``Global, regional prevalence, incidence and risk factors of knee
  osteoarthritis in population-based studies,''
\newblock {\em EClinicalMedicine}, vol. 29, pp. 100587, 2020.

\bibitem{kellgren1957radiological}
JH~Kellgren and JS~Lawrence,
\newblock ``Radiological assessment of rheumatoid arthritis,''
\newblock {\em Annals of the rheumatic diseases}, vol. 16, no. 4, pp. 485,
  1957.

\bibitem{panfilov2021deep}
E~Panfilov, A~Tiulpin, MT~Nieminen, et~al.,
\newblock ``Deep learning-based segmentation of knee mri for fully automatic
  subregional morphological assessment of cartilage tissues: Data from the
  osteoarthritis initiative,''
\newblock {\em Journal of Orthopaedic Research{\textregistered}}, 2021.

\bibitem{morales2020learning}
A~Morales~Martinez, F~Caliva, I~Flament, et~al.,
\newblock ``Learning osteoarthritis imaging biomarkers from bone surface
  spherical encoding,''
\newblock {\em Magnetic resonance in medicine}, vol. 84, no. 4, pp. 2190--2203,
  2020.

\bibitem{wang2019total}
T~Wang, K~Leung, K~Cho, et~al.,
\newblock ``Total knee replacement prediction using structural {MRIs} and {3D}
  convolutional neural networks,''
\newblock in {\em International Conference on Medical Imaging with Deep
  Learning--Extended Abstract Track}, 2019.

\bibitem{tolpadi2020deep}
AA~Tolpadi, JJ~Lee, V~Pedoia, and S~Majumdar,
\newblock ``Deep learning predicts total knee replacement from magnetic
  resonance images,''
\newblock {\em Scientific reports}, vol. 10, no. 1, pp. 1--12, 2020.

\bibitem{tiulpin2019multimodal}
A~Tiulpin, S~Klein, SMA Bierma-Zeinstra, et~al.,
\newblock ``Multimodal machine learning-based knee osteoarthritis progression
  prediction from plain radiographs and clinical data,''
\newblock {\em Scientific reports}, vol. 9, no. 1, pp. 1--11, 2019.

\bibitem{raghu2019transfusion}
M~Raghu, C~Zhang, J~Kleinberg, and S~Bengio,
\newblock ``Transfusion: Understanding transfer learning for medical imaging,''
\newblock {\em arXiv preprint arXiv:1902.07208}, 2019.

\bibitem{mustafa2021supervised}
B~Mustafa, A~Loh, J~Freyberg, et~al.,
\newblock ``Supervised transfer learning at scale for medical imaging,''
\newblock {\em arXiv preprint arXiv:2101.05913}, 2021.

\bibitem{kopuklu2019resource}
O~Kopuklu, N~Kose, A~Gunduz, and G~Rigoll,
\newblock ``Resource efficient {3D} convolutional neural networks,''
\newblock in {\em Proceedings of the IEEE/CVF International Conference on
  Computer Vision Workshops}, 2019, pp. 0--0.

\bibitem{dosovitskiy2020image}
A~Dosovitskiy, L~Beyer, A~Kolesnikov, et~al.,
\newblock ``An image is worth 16x16 words: {Transformers} for image recognition
  at scale,''
\newblock {\em arXiv preprint arXiv:2010.11929}, 2020.

\bibitem{arnab2021vivit}
A~Arnab, M~Dehghani, G~Heigold, et~al.,
\newblock ``{ViViT}: A video vision transformer,''
\newblock {\em arXiv preprint arXiv:2103.15691}, 2021.

\bibitem{xie2021cotr}
Y~Xie, J~Zhang, C~Shen, and Y~Xia,
\newblock ``{CoTr}: Efficiently bridging {CNN} and {Transformer} for {3D}
  medical image segmentation,''
\newblock {\em arXiv preprint arXiv:2103.03024}, 2021.

\bibitem{zhang2021transfuse}
Y~Zhang, H~Liu, and Q~Hu,
\newblock ``Transfuse: Fusing transformers and cnns for medical image
  segmentation,''
\newblock {\em arXiv preprint arXiv:2102.08005}, 2021.

\bibitem{he2016deep}
K~He, X~Zhang, S~Ren, and J~Sun,
\newblock ``Deep residual learning for image recognition,''
\newblock in {\em Proceedings of the IEEE conference on computer vision and
  pattern recognition}, 2016, pp. 770--778.

\bibitem{perslev2019one}
M~Perslev, EB~Dam, A~Pai, and C~Igel,
\newblock ``One network to segment them all: A general, lightweight system for
  accurate {3D} medical image segmentation,''
\newblock in {\em International Conference on Medical Image Computing and
  Computer-Assisted Intervention}. Springer, 2019, pp. 30--38.

\bibitem{tran2018closer}
D~Tran, H~Wang, L~Torresani, et~al.,
\newblock ``A closer look at spatiotemporal convolutions for action
  recognition,''
\newblock in {\em Proceedings of the IEEE conference on Computer Vision and
  Pattern Recognition}, 2018, pp. 6450--6459.

\bibitem{xie2016}
S~Xie, R~Girshick, P~Dollár, et~al.,
\newblock ``Aggregated residual transformations for deep neural networks,''
\newblock {\em arXiv preprint arXiv:1611.05431}, 2016.

\bibitem{johnson2019survey}
JM~Johnson and TM~Khoshgoftaar,
\newblock ``Survey on deep learning with class imbalance,''
\newblock {\em Journal of Big Data}, vol. 6, no. 1, pp. 1--54, 2019.

\bibitem{saito2015precision}
T~Saito and M~Rehmsmeier,
\newblock ``The precision-recall plot is more informative than the {ROC} plot
  when evaluating binary classifiers on imbalanced datasets,''
\newblock {\em PloS one}, vol. 10, no. 3, pp. e0118432, 2015.

\bibitem{zhu2021}
L~Zhu,
\newblock ``{THOP},'' \url{https://github.com/Lyken17/pytorch-OpCounter}, 2021.

\bibitem{taleb2021multimodal}
A~Taleb, C~Lippert, T~Klein, and M~Nabi,
\newblock ``Multimodal self-supervised learning for medical image analysis,''
\newblock in {\em International Conference on Information Processing in Medical
  Imaging}. Springer, 2021, pp. 661--673.

\bibitem{azizi2021big}
S~Azizi, B~Mustafa, F~Ryan, et~al.,
\newblock ``Big self-supervised models advance medical image classification,''
\newblock {\em arXiv preprint arXiv:2101.05224}, 2021.

\bibitem{namiri2021deep}
NK~Namiri, J~Lee, B~Astuto, et~al.,
\newblock ``Deep learning for large scale {MRI}-based morphological phenotyping
  of osteoarthritis,''
\newblock {\em Scientific reports}, vol. 11, no. 1, pp. 1--10, 2021.

\bibitem{casula2017elevated}
V~Casula, MJ~Nissi, J~Podlipsk{\'a}, et~al.,
\newblock ``Elevated adiabatic {T1$\rho$} and {T2$\rho$} in articular cartilage
  are associated with cartilage and bone lesions in early osteoarthritis: A
  preliminary study,''
\newblock {\em Journal of Magnetic Resonance Imaging}, vol. 46, no. 3, pp.
  678--689, 2017.

\end{thebibliography}
